\documentclass[letterpaper,useAMS,usenatbib]{mn2e}
\usepackage{graphicx,color,epsfig,natbibmnfix}

\newcommand{\be}{\begin{equation}}
\newcommand{\ee}{\end{equation}}

\newcommand{\apj}{ApJ}

\newcommand{\mnras}{MNRAS}
\newcommand{\aap}{A\&A}
\newcommand{\araa}{ARA\&A}
\newcommand{\apjl}{ApJL}
\newcommand{\aj}{AJ}

\newcommand{\icarus}{ICARUS}

\def\ltsima{$\; \buildrel < \over \sim \;$}
\def\simlt{\lower.5ex\hbox{\ltsima}}
\def\gtsima{$\; \buildrel > \over \sim \;$}
\def\simgt{\lower.5ex\hbox{\gtsima}}

\def\msun{{\,{\rm M}_\odot}}

\def\del#1{{}}

\title[Hydrogen-loosing planets in transition discs]{Hydrogen-loosing planets in
  transition discs around young protostars}

\author[Sergei Nayakshin]{Sergei Nayakshin\\ 
Department of Physics \& Astronomy,
  University of Leicester, Leicester, LE1 7RH, UK\\
}

\begin{document}

\date{Received}

\pagerange{\pageref{firstpage}--\pageref{lastpage}} \pubyear{2011}

\maketitle

\label{firstpage}

\begin{abstract}
We point out that protoplanets created in the framework of the Tidal
Downsizing (TD) theory for planet formation play a very important role for the
evolution of accretion discs hosting them. Since all TD protoplanets are
initially as massive as $\sim 10$ Jupiter masses, they are able to open very
deep gaps in their discs, and even completely isolate the inner disc flows
from the outer ones. Furthermore, in contrast to other planet formation
theories, TD protoplanets are mass donors for their protostars. One
potentially observable signature of planets being devoured by their protostars
are FU Ori like outbursts, and episodic protostar accretion more generally, as
discussed by a number of authors recently.

Here we explore another observational implication of TD hypothesis: dust poor
inner accretion flows, which we believe may be relevant to some of the
observed mm-bright transitional discs around protostars. In our model, a
massive protoplanet interrupts the flow of the outer dust-rich disc on its
protostar, and at the same time looses a part of its dust-poor envelope into
the inner disc. This then powers the observed gas-but-no-dust accretion onto
the star.  Upon a more detailed investigation, we find that this scenario is
quite natural for young massive discs but is less so for older discs, e.g.,
those whose self-gravitating phase has terminated a fraction of a Million year
or more ago.  This stems from the fact that TD protoplanets of such an age
should have contracted significantly, and so are unlikely to loose much
mass. Therefore, we conclude that either (i) the population of ``transition
discs'' with large holes and dust-poor accretion is much younger than
generally believed; or (ii) there is a poorly understood stage for late
removal of dust-poor envelopes from TD planets; (iii) another explanation for
the observations is correct.
\end{abstract}


\section{Introduction}\label{sect:intro}

In the standard paradigm for planet formation \citep[e.g.,][]{Wetherill90} and
chapters 4-6 in \cite{Armitage10}, large and massive protostellar discs
eventually evolve into much less massive ``protoplanetary'' discs. The latter
represent the end phase of star formation and the beginning phase of planet
formation. Such discs can be now observed around nearby pre-main sequence
stars \citep[see, e.g.,][for a recent review]{DM10Review}.

An especially interesting sub-class of protoplanetary discs are the so-called
``transition discs'', broadly defined as objects showing mid-IR excess above
the protostar's emission, but lacking the corresponding near-IR
excess. Near-IR emission is characteristic of the inner $\simlt$ few AU disc
region, and mid-IR emission is a signature of the outer $\sim$ tens of AU
disc. Therefore, transition discs are interpreted as protoplanetary discs with
inner holes. High-resolution sub-millimetre continuum images of several
sources, such as LkH$\alpha$ 330 \citep{BrownEtal08} and GM Tau
\citep{HughesEtal09}, indeed show sharp inner holes tens of AU in radius and
thus confirm validity of this interpretation of spectral energy distributions
(SEDs) of transition discs.

Since gas discs around protostars disappear with time, discs with inner holes
are believed to be the discs that are at the beginning of their "disc clearing
stage". Since the fraction of such systems is low, e.g., $\sim 0.1$ of the
sample, this was taken to mean that the transitional disc phase takes $\sim
10$\% of the disc lifetime, e.g., from $\sim 10^5$ yrs to a few times that
\citep{AndrewsWilliams05,AndrewsWilliams07}. Such a behaviour naturally appears
in models in which the disc is first evolving on a long viscous time scale and
is then photo-evaporated rapidly when the accretion rate drops to low $\sim
10^{-10} \msun/$yr values \citep{ClarkeEtal01a,AlexanderEtal06}.

It then came as a considerable surprise that transition discs with large $R >
15$ AU to $\sim 70$ AU holes are quite common among {\em mm-bright discs}:
recent observations of \cite{AndrewsEtal11} show that these discs comprise
between $1/5$ to $\sim 2/3$ of their sample of such objects.  These massive
discs should not really be the ones in the process of dispersal yet. The disc
photo-evaporation models cannot easily create such large inner holes for the
high accretion rates observed \citep[see][]{OwenEtal12}. Therefore,
\cite{OwenClarke12} suggest that " ... either (i) ``transition discs'' are a
misnomer and these objects do not physically represent an evolutionary stage
from disc-bearing and disc-less systems; or (ii) discs with inner holes can
have multiple physical origins and the two populations shown in the data
indicate two distinct classes of objects."

The consensus seems to be building up to blame one or several massive planets
per parent star, rather than disc photo-evaporation, for creating the observed
gaps in the transition discs with large gaps
\citep{AndrewsEtal11,ZhuEtal12b}. Physically, a giant planet of a few Jupiter
mass or heavier pushes the gas away radially from its location by applying
gravitational torques on the disc \citep{LinPap86}. This interaction first
opens a wide gap in the protoplanetary disc. The inner disc then drains onto
the protostar while the outer is held off by the planet; this results in a
completely evacuated inner hole in the disc \citep{RiceEtal03b}, potentially
explaining the observations of large holes in transition discs.

However, many of the transition discs with inner holes have large {\em gas}
accretion rates onto the protostar, e.g., up to $\sim 10^{-8}\msun$/yr,
requiring that only the dust flow was interrupted by the planet but not that
of the gas \citep{AndrewsEtal11}.  One plausible way out of this is a
  combination of dust filtration on gap edges and dust growth in the inner
  disc \citep{ZhuEtal12b}. In particular, \cite{RiceEtal06} showed that since
  gas pressure gradient at the outer edge of a gap opened by a massive planet
  is positive, rather than negative in a disc without a gap, grain particles
  migrate outward in the frame comoving with the gas. This contrasts with the
  usual situation when dust grains migrate inward. The outward migration speed
  depends on the size of a particle and the properties of the gap, but
  generally large grain particles, e.g., those larger than $\sim 1$ mm migrate
  outward more rapidly than gas moves inward due to viscous torques. Therefore
  such particles are stopped effectively at the gap from penetrating the inner
  disc. However, smaller particles \citep[smaller than about 10 $\mu$m,
    see][]{RiceEtal06} are so tightly bound to the gas by aerodynamical
  friction that they must follow the gas motion. These particles would follow
  the gas into the inner disc, through the gap, and would then still be
  observable in NIR, in contradiction to the observations
  \citep{AndrewsEtal11}. To solve this difficulty, \cite{ZhuEtal12b} propose
  that small dust particles that did penetrate the gap grow to large sizes, so
  that their opacity drops to barely detectable levels.

In this paper we propose a completely different take on the problem of
dust-poor accretion onto the protostars, based on a crucial element of the new
``Tidal Downsizing'' scenario \citep[TD;][]{BoleyEtal10,Nayakshin10c} for
planet formation -- {\em mass loss from the giant protoplanets}. This mass
loss process is unique to TD scenario. The two other models for forming giant
planets -- the Core Accretion (CA) and the Gravitation disc Instability (GI)
\citep[see, e.g.,][]{Boley09}, assume that planets always {\em grow} in mass
by accretion of gas from the disc as they age. In both of these models the
inner dust-poor disc must be a continuation of the larger scale accretion
disc, perhaps modified by the presence of the planet as discussed above. In TD
model, however, most of the protoplanets, having been formed in the outer cold
disc at $R\sim 100$ AU, give their gaseous envelopes back to the {\em inner
  disc} when they migrate sufficiently close to the parent star \citep[see
  simulations by][]{ChaNayakshin11a}.

In this paper we show that (i) TD planets, being initially more massive than
CA planets, are able to open complete gaps in the protoplanetary
discs. Therefore, in our model for transition discs with large inner cavities,
the outer dust-rich accretion flow, in both gas and dust, is completely
disconnected from the inner region. (ii) TD planets must loose their envelopes
on the way to becoming the present day planets. The accretion flow inward of
the gap can be "restarted" when these envelopes are accreted by the protostar
\citep{NayakshinLodato12}. In this picture, the protoplanet behaves like a
lower mass version of the mass-loosing secondary in a compact stellar binary
system with mass transfer \citep[e.g.,][]{Ritter88}. (iii) Crucially, the
envelopes of TD protoplanets are dust-poor \citep{Nayakshin10c} because the
dust is expected to sediment to their centres. Therefore, we propose that the
inner accretion flow of dust-poor transition discs is nothing less than the
planet's envelope devouring stage of Tidal Downsizing for one of the inner
planets.

Below we explore this model in greater detail.

\section{Planets and protoplanetary discs}\label{sec:connection}

\subsection{TD model}

The first suggestion that planets are initially much more massive than they
presently are, was made, to the best of our knowledge, by \cite{Kuiper51b},
who proposed that planets grew from self-gravitating condensations in the
Solar Nebula. His Figure 1 shows that he considered it quite possible that
protoplanets were initially much more massive. Specifically, on page 9 on his
manuscript he suggests that each of the Solar System planet was formed from
$\sim$0.003 Solar masses of gas and dust, a guess that, amazingly, is
consistent with the TD model within a factor of a few. \cite{McCreaWilliams65}
added practical detail to this scenario, explaining how the differentiation of
materials within such proto-planets could have taken place. They showed that
microscopic grains grow and sediment to the centre of such gas clumps within a
few thousand years \citep[cf. also][]{Boss97}. Proto-Earth formation in this
picture is complete when the gas envelope of the proto-planet is removed by
Solar tides.

However, since the process of planet migration was unknown until later
\citep{LinPap79,GoldreichTremaine80}, a physically complete and
self-consistent scenario for this {\em top-down} scenario for planet formation
was not found, and the model was essentially given up \citep{DW75} until very
recently \citep{BoleyEtal10,Nayakshin10c}. The modern version of the top-down
scenario is essentially Gravitational disc Instability model for planet
formation upgraded by the physics of giant planet migration and disruption. In
the TD scenario, massive $\sim 10 M_J$
\citep[e.g.,][]{BoleyEtal10,ForganRice11} gaseous clumps are formed in the
outer disc by the disc's GI. The clumps then migrate inward, while dust
sediments to the centre of the clumps
\citep{HS08,HB11,Nayakshin10a,Nayakshin10b}. A partial removal of their gas
envelopes by gravitational tides, irradiation or other effects
\citep{BossEtal02,Nayakshin10c,NayakshinCha12} results in formation of
metal-enriched giant planets such as Jupiter or Saturn, whereas a complete
removal of the gas results in terrestrial like planets.

TD model has direct connections to protostellar disc evolution. A story line
very similar to TD scenario, but minus the role of dust grains, has been
developing in the field of protostellar evolution since \cite{VB06}. These
authors found that massive gaseous clumps formed by GI in the outer disc
migrate rapidly inward and get "accreted" by the star in bursts \citep[see
  also][]{VB10}. Such accretion bursts appear to be a natural explanation for
the FU Ori outbursts of young protostars where the star accretes mass at rates
as high as $\sim 10^{-4} \msun$~yr$^{-1}$ \citep{HK96}. A detailed
investigation of this episodic accretion model for the FU Ori outbursts by
\cite{NayakshinLodato12} confirmed its potential promise, and also showed that
the inner disc behaviour may be additionally modulated by the thermal
ionisation instability. This implies that both episodic accretion and thermal
ionisation instability \citep{Bell94} may play a role in explaining the
observed FU Ori outbursts.  The other side of this bursty accretion picture --
long periods of relatively low protostellar accretion rates -- may naturally
explain \citep{DunhamVorobyov12} the "luminosity problem" of young stars
\citep{HK96}.

\subsection{Protoplanetary disc sculpting by TD planets}

A sufficiently massive protoplanet may open a gap in the accretion disc and
thus regulate the rate of mass flow onto the protostar
\citep{LinPap86}. \cite{CridaEtal06} show with 2D simulations that the inner
disc surface density is decreased by a factor of 10 (compared with a disc with
no embedded planet) when the parameter ${\cal P} \equiv 3H/4R_H + 50/ ({\cal
  R} q) \approx 1$, where $q = M_p/M_*$, $R_H$ is the Roche lobe radius, $R_H
= R (q/3)^{1/3}$, ${\cal R} = Rv_K/\nu$ is the Reynolds number, which can be
re-written as ${\cal R} \approx (R/H)^2\alpha^{-1}$ for the \cite{Shakura73}
disc model with viscosity parameter $\alpha$ and the disc vertical aspect
ratio $H/R$. For $m_1 = M_p/(10 M_J)$, and $H/R = 0.05$, we have
\begin{equation}
{\cal P} \approx {1\over 4 m_1^{1/3}} +{\alpha_{2}\over 8 m_1} \;,
\label{P}
\end{equation}
where $\alpha_{2}= \alpha/0.01$. Furthermore, the smaller the value of ${\cal
  P}$, the deeper the gap (the stronger the suppression of the inner
disc). For $m_1 = 1$, for example, ${\cal P} \approx 3/8$. Such a massive
planet opens an infinitely deep gap in the disc according to equation (13) of
\cite{CridaMorbidelli07}, that is, the inner disc is completely cut off from
the outer one by the planet. Equation \ref{P} predicts, on the other hand,
that ${\cal P} \approx 1$ for $M_p = 2 M_J$, thus the gap is expected to be
leaky, so that $\sim 10\%$ of gas is able to filter through to the inner disc
\citep{CridaEtal06}. In accord with these results, \cite{NayakshinLodato12}
found that protoplanetary discs with $m_1 \sim 1$ protoplanets feature
infinitely deep gaps, so that the inner disc empties out on the protoplanet
and becomes devoid of any gas flow for a period of time.

Given these results, protoplanets are much more important for the evolution of
protoplanetary discs in TD scenario than they are for discs with CA-built
planets. Indeed, in TD scenario, {\em every} protoplanet, including those that
go on to eventually make "tiny" planets such as the Earth is initially quite
massive, e.g., from a few to over 10 $M_J$. In other words, {\em essentially
  every} planet before the downsizing step may be able to open a very deep gap
in the TD scenario. Thus we expect discs with empty or very strongly depressed
inner flows to be a much more frequent feature in the TD picture. If we accept
that FU Ori outbursts are produced by giant protoplanets being torn into
pieces and then accreted by their protostars, and that this happens $\sim
10-20$ times during assembly of a typical star \citep{HK96}, we would expect
that discs with inner holes/strong depressions should exist for a good
fraction, e.g., one comparable to unity, of all mm-bright (that is massive)
protoplanetary discs.

In addition to this, TD scenario planets not only open the gaps and thus hold
off accretion of gas and dust beyond their orbits, they can also feed their
protostars with a significant amount of matter, given their masses. To
appreciate this, we note that the total mass of all the Solar System
protoplanets in the CA framework is about 1.5 Jupiter masses, while in TD
picture this is somewhere between 40 and 80 Jupiter masses (assuming that
Solar System planets were built from gas embryos of mass between 5 and 10
Jupiter masses).

With this in mind, we echo the suggestion of \cite{OwenClarke12} that some
transitions discs may not represent systems where the discs are being
removed. We suggest that the flow of matter in these discs, both in gas and
dust, is temporarily interrupted by very deep gaps. While the outer disc is
held off by one or more embedded planets \citep[as suggested earlier by a
  number of authors, e.g.,][]{RiceEtal06,ZhuEtal12b}, the inner disc is fed by
dust-poor gas lost by one or more planets in the gap.

In the simulations of \cite{NayakshinLodato12}, the period of the completely
empty inner disc (zero accretion rate onto the protostar) ends when the
protoplanet is pushed deep enough so that it fills its Roche lobe and starts
loosing the mass through the inner Lagrangian L1 point. The inner disc is then
quickly re-filled {\em by the material lost by the planet} and accretes onto
the protostar at very high accretion rate, leading to an FU Ori-like
outburst. These simulations confirm the earlier suggestions by
\cite{VB06,BoleyEtal10} that tidal disruption of giant protoplanets may
produce accretion outbursts with properties required to explain the FU Ori
outbursts.

Below we consider whether a physically similar model may explain the
Transition discs with {\em large} inner dust cavities. {\bf To avoid potential
  confusion, we point out that the outer self-gravitating portions of
  relatively massive protoplanetary discs may have larger $H/R$ than the value
  of 0.05 used in equation (1) above, so that the protoplanets manage to
  migrate in type-I regime there
  \citep[e.g.,][]{BaruteauEtal11,BoleyEtal10,ChaNayakshin11a} Although the
  transition in the migration regime of such massive protoplanets is a subject
  of ongoing research (S.-J. Paardekooper, private communication), the
  transition is generally expected to occur where the disc mass interior of
  the planet becomes comparable to that of the protoplanet, so that type-I
  migration rate slows down \citep[see also the physically related case of a
    binary super-massive black hole migration in][]{LodatoEtal09}}.

\section{Steady state mass loss from protoplanets}\label{sec:loss}

\subsection{Terminology}

We shall first consider steady-state scenarios -- and will find them unlikely
-- in which the inner dust-poor accretion disc is in a qasi-steady state, so
that the mass is deposited into the disc by the protoplanet at the same rate
as it is accreted by the protostar. An example of such a situation is
quasi-equilibria found in some of the \cite{NayakshinLodato12} simulations.

In TD scenario, giant protoplanets go through two stages on the way to
becoming fully fledged dense giant gas planets
\citep{BoleyEtal10,Nayakshin10a,Nayakshin11b}. The first stage begins with
formation of the protoplanet in the outer cold disc by gravitational
instability. The ``first protoplanet'' is a rather fluffy gas clump in which
hydrogen is molecular; the clump's central temperature is $\sim$ hundreds of
Kelvin and its radial extent is a few to ten AU. These clumps cool rapidly:
within some $10^3$ to at most $10^5$ yrs \citep[depending on the clump's mass
  and grain opacity, see][]{HB11} the central temperature approaches $T\approx
2000$ K at which point hydrogen molecules dissociate rapidly, and the clump
collapses hydrodynamically to sizes of the order of $0.03$ AU or even smaller
\citep[cf. \S 3.4 in][]{NayakshinLodato12}. This collapse begins the second
stage in the prtoplanet evolution, when most of hydrogen is atomic or
partially ionised.  This picture is very similar to the evolution from the
first to second cores in star formation \citep{Larson69}, although here mass
of the protoplanet may remain fixed during contraction and collapse. We shall
thus call protoplanets dominated by atomic or inosed hydrogen ``second
planets'' below for brevity.

\subsection{Roche lobe overflow mass loss}\label{sec:roche}

Mass transfer from a Roche lobe-filling secondary to the primary has been a
subject of numerous papers in the context of cataclysmic binaries
\citep[e.g.,][]{Ritter88}. The secondary's mass loss rate is found to be a
very strong function of $\Delta r = r_p - r_H$ where $r_p$ is the planet's
radius. Before the Roche lobe of the secondary is filled, $\Delta r < 0$, the
mass transfer rate has an exponential form reflecting the exponential decrease
of density with height in the stellar atmosphere \citep[cf. eq. 9
  of][]{Ritter88}:
\begin{equation}
{d M_p\over dt} = - {\rho_{\rm ph} c_{\rm ph} r_H h_p \over e^{1/2}} \exp\left[ 
  {\Delta r \over h_p} \right]\;,
\label{dotM1}
\end{equation}
where $M_p$ is the planet's mass, $h_p = k T_{\rm ph}r_p^2/\mu G M_p$ is the
scale height of the planet's atmosphere, $\mu$ is the mean molecular weight,
here taken to be $2.3 m_p$, $T_{\rm ph}$, $\rho_{\rm ph}$ and $c_{\rm ph}$ are
the photosphere's density and sound speed, respectively. The Hill's radius is
given by
\begin{equation}
r_H = a (M_p/3M_*)^{1/3}\;,
\label{rh}
\end{equation}
where $a$ is the planet-star separation. $r_H$ is approximately equal to the
Roche lobe radius.

Here we are interested in planets situated at $a\simgt 10$
AU. \cite{Nayakshin10a}, using a simple $\kappa(T)\propto T$ dust opacity
model for first protoplanets, obtains that at large time $t$ since the formation of the clump, the
clump's radius is $r_p \approx 0.8$~AU~$ (10^4$~yr$/t)^{1/2}$, 
independently of the clump's mass. Given the planet's luminosity \citep[see \S
  3 in][]{Nayakshin10b}, we obtain planet's photosphere -- effective --
temperature of about 100 K. The photosphere's scale height is then very small
compared to $r_p$,
\begin{equation}
{h_p \over r_p} = {k T_{\rm ph} r_p \over G M_p \mu} \approx 0.03\;.
\label{hp}
\end{equation}
This shows that unless $r_p$ is almost exactly equal to $r_H$, the factor
$\exp (\Delta r/ h_p)$ is very small.  Speaking qualitatively, the Roche lobe
overflow rate is nearly zero when $r_p < r_H$ and then suddenly becomes very
large (up to $\sim 10^{-4}\msun$~yr$^{-1}$) when $r_p > r_H$
\citep{NayakshinLodato12}. It would thus be a surprising coincidence to have
the planet to loose mass at the observed gas accretion rate range
($10^{-10}-10^{-8}\msun$~yr$^{-1}$).

 The arguments presented just above are for the first protoplanets. As we
 mentioned above, massive $\sim 10 M_J$ H$_2$ clumps survive only for $\sim$
 (few to few tens) of thousand years before collapsing to much denser second
 configuration. This is too short given that transition discs probably exist
 for $\sim 10^5$ years: finding a protoplanet in the first stage inside the
 transition disc cavity should be unlikely. The protoplanet may be inflated by
 tidal torques \citep{GoldreichSoter66} if its orbit is eccentric, but the
 relevant time scales \citep[e.g., \S 3.2 in][]{BodenheimerEtal03} are longer
 than the Hubble time at $a\simgt 10$~AU.

The second stage protoplanets are much more compact than the first stage ones,
so that the ratio $h_p/r_p$ is yet smaller for them. We conclude that a
quasi-steady state tidal disruption (Roche lobe overflow) of giant
protoplanets is unlikely to match the observed protostellar accretion rates of
transition disc systems.

\subsection{Photo-evaporation of the second stage planets}\label{sec:2nd}

Alternatively, planets may loose mass by photo-evaporation
\citep{Lecavelier07}. However, \cite{OwenJackson12} showed that mass loss of
the  second stage planets due to photo-evaporation does not
exceed $\sim 10^{-10}\msun$~yr$^{-1}$ even for planets located at distances as
close as $0.1$ AU from the star. The weight of the argument is best
appreciated by considering the energy-limited photo-evaporation, which is
essentially the maximum possible outflow rate in which all the ionising power
goes into driving the outflow, with no re-radiation losses \citep[see,
  e.g.,][]{OwenJackson12}. The fraction of stellar radiation intercepted by
the planet is $\zeta_{\rm rad} = \pi r_p^2/4\pi a^2$, which is only $\sim
0.0025$ for a few $M_J$ planet completely filling its Roche lobe. The
combined X-ray and UV luminosity of a typical T-Tauri star is $L_{\rm
  ion}\sim$ a few $\times 10^{30}$ erg~s$^{-1}$. This yields the maximum
outflow rate
\begin{equation}
\dot M_{\rm en} \sim {\zeta_{\rm rad} L_{\rm ion}\over c_{\rm ph}^2} \simeq
10^{-10} \msun\hbox{ yr}^{-1}\;,
\end{equation}
where $c_{\rm ph}\sim 10$~km~s$^{-1}$ is the sound speed in the evaporated
outflow. Since $r_p$ is only a few percent of $r_H$ for the second stage
planets at $a \sim 10$ AU, we conclude that the photo-evaporative mass loss
rate is completely inadequate to explain the observed gas accretion rates of
up to $10^{-8}\msun$~yr$^{-1}$ in the \cite{AndrewsEtal11} sample.

\subsection{Circum-planetary disc and its removal}

Recently, \cite{CalvagniEtal12} demonstrated the importance of rotation for
clumps formed in the outer gravitationally unstable disc during the transition
from the first to the second protoplanetary stages. In particular, when H$_2$
dissociation occurs only $\sim 50$\% of mass of the initial clump collapses
into a quasi-spherical second planet, with the rest orbiting the collapsed
part in a thickish ($H/R \sim 0.5$) rotationally supported circum-planetary
disc.  This disc (instead of the planet's atmosphere) could in principle play
the role of the dust-poor mass reservoir being stripped away and consumed by
the parent star.

Circum-planetary discs are truncated by tidal torques from the star at around
$r_{\rm d} \sim 0.3 r_H$ to $0.4 r_H$
\citep{AyliffeBate09,MartinLubow11}. Thus their outer edges are still quite
far from the L1 point. The most likely way to loose these discs is by
photo-evaporation \citep{MitchellStewart11}. However, applying the equation A7
in \cite{AdamsEtal04} for the present case, the rate at which the discs are
likely to be photo-evaporating is modest, $ \dot M_d \le 10^{-10}
\msun$~yr$^{-1}$. Furthermore, this is likely to be an over-estimate, since we
find that H$_2$ molecule densities in the evaporative flows would be $\sim
10^9$~cm$^{-3}$, and therefore the H$_2$ gas should self-shield itself from
the FUV radiation \citep[see the end of \S 2.1 and \S 2.3 in][]{AdamsEtal04},
further reducing the estimate. Finally, the path of the protostar's
photo-ionising radiation to the planet is likely to be blocked by the observed
inner gas accretion flows since the inner disc aspect ratio ($H/R\sim 0.1$)
may be larger than $r_p/a$. Therefore, while it appears clear that
circum-planetary gas discs can be efficiently removed by photo-evaporation on
long time scales \citep{MitchellStewart11}, it seems that their steady-state
mass loss is insufficiently large to explain protostellar accretion rates of up to
$10^{-8}\msun$~yr$^{-1}$.

\section{Non-steady scenarios}

The accretion disc viscous time at radius $R$ is
\begin{equation}
t_{\rm visc} \sim \alpha^{-1} {R^2 \over H^2} \Omega(R)^{-1} = 5\times 10^4 \hbox{ yr }
\alpha_{-2}^{-1} {R^2 \over 100 H^2} R_1^{-3/2}\;,
\label{tvisc}
\end{equation}
where $R_1 = R/10$~AU, $\alpha$ is the \cite{Shakura73} viscosity parameter,
$H/R$ is the disc geometrical aspect ratio, and $\Omega(R) = (GM_*/R^3)^{1/2}$
is the Keplerian angular frequency. The accretion disc viscous time is
comparable with the estimated lifetime of the transition disc phase,
suggesting that the disc may not necessarily be in a quasi-steady state. This
in turn implies that the observed proto-stellar accretion rates do not have to
be equal to {\em instantaneous} mass loss rates from the planet(s) in the scenario we propose.

Suppose that a massive gas planet has lost a $\Delta M \sim 1 M_J$ of its dust-poor
envelope. The average accretion rate onto the protostar as the result of this
is
\begin{equation}
<\dot M>\quad  \sim \quad {\Delta M \over t_{\rm visc}} = 10^{-8}\;
{\msun\over \hbox{yr}} t_5^{-1}\;,
\end{equation}
where $t_5 = t_{\rm visc}/(10^5$~yr). One expects that initially the accretion
rate is higher than this estimate but then decays to a somewhat lower
value. Statistically speaking, one is more likely to catch the longer decaying
part of this accretion episode, so that the range of the observed accretion
rates in the \cite{AndrewsEtal11} sample appears qualitatively natural in this
picture.

To test these ideas, we have used the code of \cite{NayakshinLodato12} to
investigate the disc-planet system evolution in the case of a partial envelope
or circum-planetary disc removal from a massive planet. In this calculation we
assume that the planet opened a gap in the accretion disc and that the inner
disc has drained onto the star before the calculation begins. We then assume
that the planet quickly looses a set amount of its envelope through the L1
point into the inner disc and then simulate the longer term evolution of the
disc-planet system.

Figure 1 presents one such calculation in which the initial mass and position
of the planet is $M_p = 12 M_J$ and $a = 40$ AU, respectively. The initial disc
 surface density profile follows the form
\begin{equation}
\Sigma_0(R) = {A_m \over R} \;\exp\left[-{R\over R_0}\right]\;\hbox{ if } R > R_{\rm hole}(0)
\label{sigma0}
\end{equation}
and $\Sigma_0(R)=0$ for $R < R_{\rm hole}(0) = 2 a$, where $R_0 = 60$ AU. The
functional form used in equation \ref{sigma0} is the same as that used by
\cite{AndrewsEtal11} to fit their observations. The normalisation constant
$A_m$ is set by requiring the total disc mass to be $M_d = 0.03 \msun$.

The planet's mass loss in this calculation is assumed to be constant during
the first $t_0=10^4$~yrs, $dM_p/dt = - \Delta M_p/t_0$, where $\Delta M_p = 3
M_J$, and then $dM_p/dt = 0$ for $t> t_0$. This mimics an essentially
instantaneous mass loss from the planet and its injection into the inner disc.

Initially the mass lost by the planet circularises in a ring at $R = a(1-r_h/a)^4$
\citep[cf. equation 38 in][]{NayakshinLodato12}, just inward of the planet's
location. It then spreads viscously in both directions. The inward facing
front of this vspreading ring then starts accreting onto the star,
whereas the outer edge starts to exchange the angular momentum with the
planet. The strong gravitational torques in this model keep the inner
dust-poor accretion flow separated from the outer dust-rich disc.

Figure 1a presents snapshots of the disc surface density profile at four
different times as indicated in the caption. {\bf We emphasise that the accretion 
flow inward of the gap (planet) is the flow of matter dumped by the planet into the 
inner disc, so it is physically distinct from the outer dust-rich disc flow.} As the 
planet is pushed closer to
the star, the outer disc edges closer as well. The inner disc shrinks in size
and mass with time. By the last snapshot shown in the figure (dotted curve),
almost all of the planet's envelope has been accreted by the star. Shortly
thereafter, the inner $\sim 10$ AU feature a complete disc hole -- in gas and
dust at the same time.

Figure 1b shows the planet-star separation evolution (solid curve) and the
dust hole radius evolution with time (dotted). Fig. 1c shows the accretion
rate evolution onto the protostar. As expected from our analytical estimate
above, the accretion rate onto the protostar indeed spans the range from 0 to
$\sim 10^{-8}\msun$~yr$^{-1}$. Note that the dust hole size and the
protostellar accretion rate are consistent in bulk with the observed mm-bright
sample of \cite{AndrewsEtal11} for $\sim 0.3$ Myr, which appears to be long
enough to account for the significant frequency of the discs with large holes.

\begin{figure}
\centerline{\psfig{file=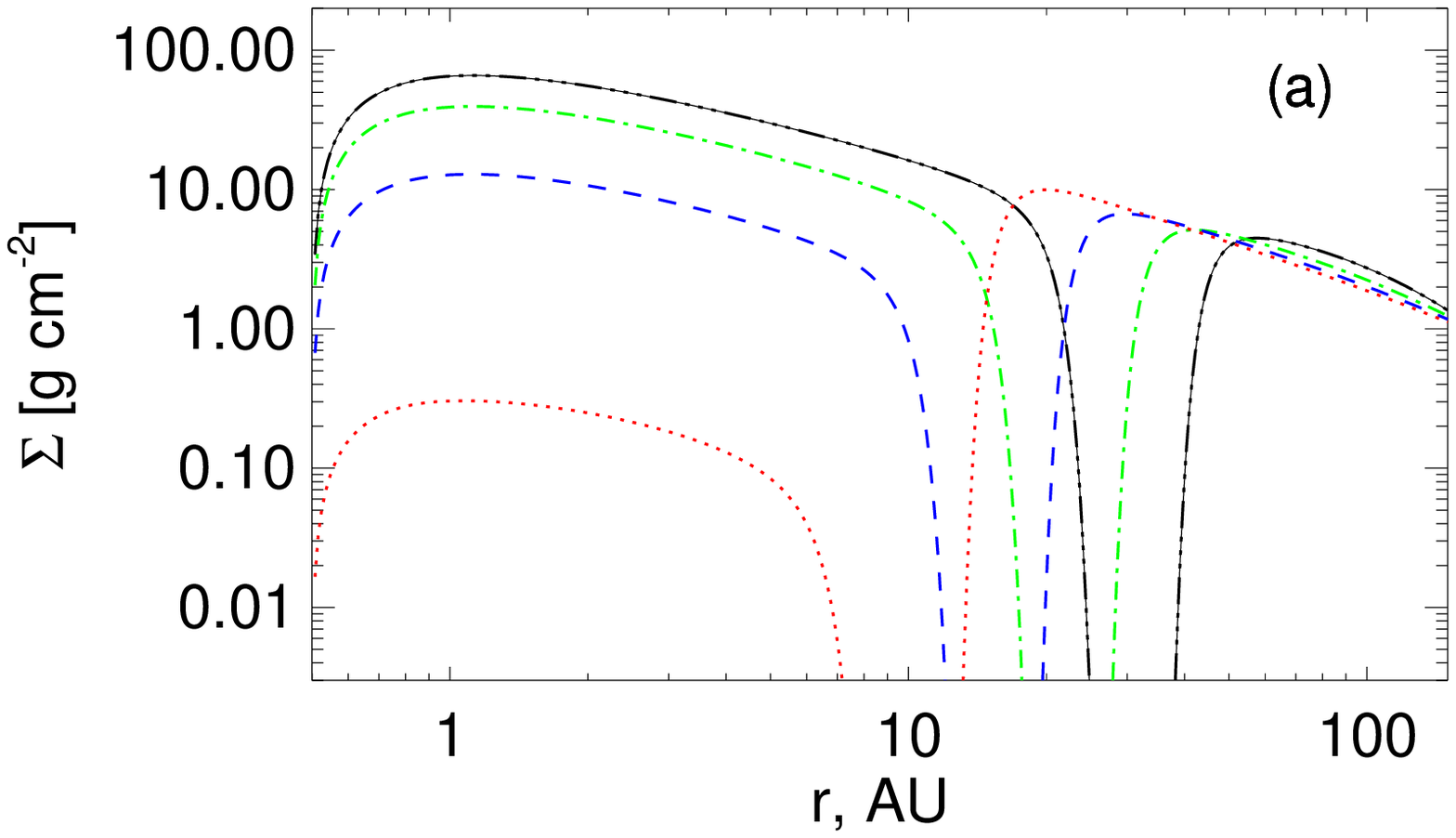,width=0.5\textwidth,angle=0}}
\vskip  -0.7 cm
\centerline{\psfig{file=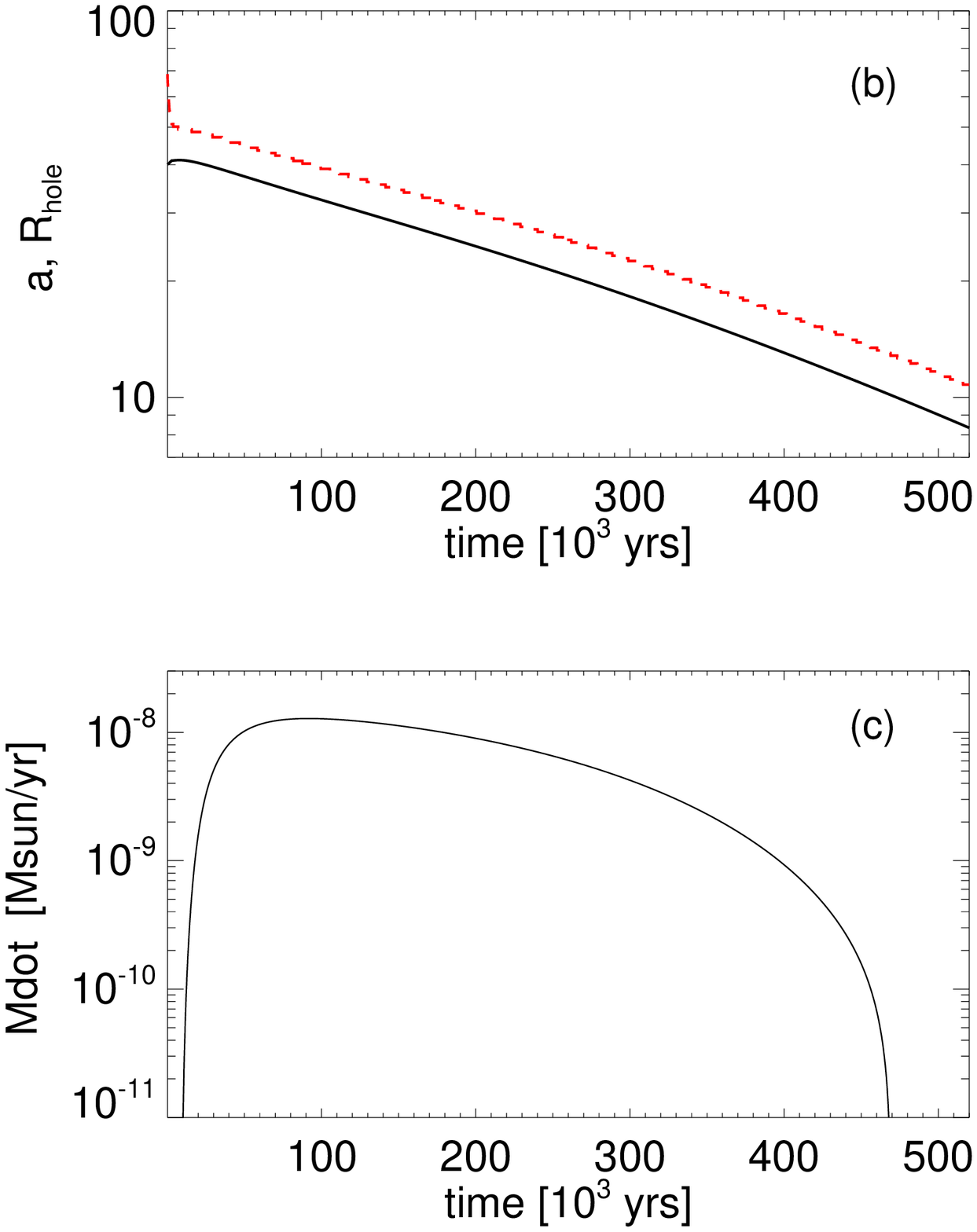,width=0.5\textwidth,angle=0}}
\caption{(a) Disc surface density profile $\Sigma(R)$ at times $t=$ 0.11,
  0.23, 0.35 and 0.46 Myr shown with the solid, dash-dotted, dashed and dotted
  curves, respectively. We emphasise that the material filling in the
    disc inward of the gap has been deposited there by the planet that lost
    its dust-poor envelope. The inner disc is physically distinct and
    separated by the gap from the outer disc. (b) The planet-star separation,
  $a$, and the size of the dust hole, $R_{\rm hole}$, versus time, shown with
  the solid and dashed curves, respectively. (c) The protostellar accretion
  rate versus time.}
\label{fig:CD_hist}
\end{figure}

\section{Discussion}

In this paper we argued that protoplanets created in the framework of Tidal
Downsizing scenario play a very important role for their protoplanetary discs,
one much more important than do planets in the CA framework. Since a typical
planet is initially as massive as $\sim 10$ Jupiter masses, TD scenario
predicts that discs with deep gaps opened by gravitational torques of the
planets must be much more frequent than in the CA picture. Furthermore, as
shown by \cite{NayakshinLodato12}, such massive planet can (temporarily)
completely stop the flow of gas and dust from the outer disc, which is not
possible for any but only the most massive CA planets. Finally, TD planets may
themselves feed inner accretion flows when their envelopes are removed by
tides or other effects.

We then suggested that the current conundrum of the observed
mm-bright transition discs with large {\em dust but not gas} holes is resolved
by combining the de-facto most popular scenario for the hole opening --
massive planet(s) inside the hole -- with the TD ideas, e.g., that massive
gaseous protoplanets have their dust sedimented and locked into the cores and
their dust-poor envelopes removed by tides or other effects. We then found
that steady-state mass loss from either ``first planets'' (molecular H$_2$) or
``second planets'' (atomic or ionised $H$) at the requisite rates of up to
$\sim 10^{-8}\msun$~yr$^{-1}$ is unlikely (\S 3). In brief, the first
planets should cool too rapidly so are somewhat unlikely to be found in
transition discs that are at least a fraction of a Myr old; the second planets
are too compact to loose mass at the observed high rates.

However, in \S 4 we pointed out that the inner dust-poor flows inside the
planet's orbit do not have to be steady state on the disc viscous time scale, 
which is of the order of $\sim 0.1$~Myr at the observed dust edges (equation
\ref{tvisc}). In particular, it appears possible that the observed accretion
rates onto the protostars in the transition discs are the residual flows
remaining after a few $M_J$ of dust-poor material was dumped by a planet into
the inner disc. One interesting possibility that we shall pursue in a future
paper is that such an episodic envelope removal from the planet is a
consequence of a much larger X-ray and ionising flux from the protostar during
the ``last'' FU Ori outburst of the protostar (the average steady-state fluxes
do not result in large enough steady-state planet photo-evaporation, see \S
3). 

One issue that requires future work is the exact mechanism through which
  protoplanets may loose the required amount of mass, which was simply
  postulated in our numerical calculation above. It is definitely easier for
  first stage planets to loose mass than it is for the more compact second
  stage ones. The first stage planets may however not live long enough for
  most of class II protostars. It may be possible that some of these are
  younger than they appear due to swelling during FU Ori outbursts
  \citep{BaraffeEtal12}. Also, the first stage protoplanets may be supported
  against contraction due to internal energy release in the manner discussed
  in \cite{NayakshinCha12}, and so may survive for longer than calculations
  neglecting the energy release at the core suggest
  \citep[e.g.,][]{Nayakshin10c,HB11}.

Furthermore, there are other plausible ways in which the necessary dust-poor
material can be lost by the planets inside the hole at the requested rates.
If there is more than one massive planet inside the hole, their mutual
interactions may drive secular evolution of the system, pumping planets'
eccentricities and/or changing their inclinations. One of the planets (or the
circum-planetary disc around it) may then loose mass at each passage of the
pericentre. The matter lost by the planet would then accrete onto the star on
much longer viscous time scales in a seemingly steady-state accretion.

We note that we did not consider here the photo-evaporation of the disc
itself, a process which may influence the structure of the disc in additional
ways \citep[e.g.,][]{RosottiEtal13}.

\section{Conclusions}

We suggested that the process of dust sedimentation invoked in the Tidal
Downsizing scenario for planet formation {\bf potentially} provides an attractive explanation
for the origin of the dust-poor accretion flows onto protostars in mm-bright
transition disc systems. Since dust grows and sediments to the centre of TD 
protoplanets, their envelopes are dust-poor. To explain the fact that most planets are
far less massive than the initial protoplanet mass of many Jupiter masses, these envelopes
must be lost to the parent stars. It would thus be natural to find at least some protostars
to accrete dust-poor matter.  The
expected accretion rate of protoplanet's envelopes onto the protostar in this
picture is, within an order of magnitude, $\dot M \sim$ a Jupiter mass divided
by $\sim 10^5-10^6$~yr (the order of magnitude disc viscous time and also the
estimated age of the transition discs). This predicts a typical protostar
accretion rate $\dot M \sim 10^{-9}\msun$~yr$^{-1}$, which is logarithmically
in the middle of the observed accretion rates \citep{AndrewsEtal11}.

{\bf However, one difficulty of our model is in the required age of the
  protoplanets. {\em If} the observed transitional discs are indeed $\sim 1$
  Myr past the massive self-gravitating stage of the disc evolution, then the
  protoplanets embedded in them would have to be similarly old. At this "old"
  age they should have contracted to very high densities; it appears very hard
  for them to loose much mass by any of the mechanisms we considered in this
  paper. A resolution to this may be found if we over-estimate the age of the
  discs in these systems; we under-estimate the planet mass loss at a fraction
  of a Myr; or this model does not apply to the older of the observed
  systems.}

\section{Acknowledgments}

Theoretical astrophysics research in Leicester is supported by an STFC Rolling
Grant. The author acknowledges a number of useful discussions with Cathie
Clarke and Richard Alexander on the photo-evaporation rates of planes.


\label{lastpage}

\end{document}